\documentclass[aps,twocolumn,showpacs,floatfix]{revtex4}

\usepackage{amssymb}
\usepackage{amsmath}
\usepackage{epsfig}
\usepackage{subfigure}

\begin{document}

\title[Constraints on the Variation of G from Primordial
  Nucleosynthesis]{Constraints on the Variation of G from Primordial Nucleosynthesis}

\author{Timothy Clifton}
\email{T.Clifton@damtp.cam.ac.uk}
\affiliation{DAMTP, Centre for Mathematical Sciences, University of Cambridge,
Wilberforce Road, Cambridge, CB3 0WA, UK}
\author{Robert J. Scherrer}
\email{Robert.J.Scherrer@Vanderbilt.Edu}
\affiliation{Department of Physics and Astronomy, Vanderbilt
  University, 6301 Stevenson Center, Nashville, TN, 37235, USA}
\author{John D. Barrow}
\email{J.D.Barrow@damtp.cam.ac.uk}
\affiliation{DAMTP, Centre for Mathematical Sciences, University of Cambridge,
Wilberforce Road, Cambridge, CB3 0WA, UK}
\date{\today }
\pacs{04.50.+h, 26.35.+c, 98.80.Ft, 98.80.Jk}

\begin{abstract}
We study here the effect of a varying $G$ on the evolution of the
early Universe and, in particular, on primordial nucleosynthesis.  This
variation of $G$ is modelled using the Brans--Dicke theory as well as
a more general class of scalar--tensor theories.  Modified
nucleosynthesis codes are used to investigate this effect and the
results obtained are used to constrain the parameters of the theories.  We extend
previous studies of primordial nucleosynthesis in scalar--tensor theories by
including effects which can cause a slow variation of $G$ during
radiation domination and by including a late--time accelerating phase
to the Universe's history.  We include a brief discussion on the epoch of
matter--radiation equality in Brans--Dicke theory, which is also of
interest for determining the positions of the cosmic microwave background power--spectrum peaks. 
\end{abstract}

\maketitle

\section{Introduction}

The possibility of the physical ``constants'' taking different values at
different times in the Universe's history has recently received much
attention with the apparent observation that the fine
structure constant had a different value in the distant past
\cite{Web99}.  These observations, if correctly interpreted \cite{footnote},
motivate further study of the effects of any variation on other
physical constants.  The scalar--tensor (ST) theories of gravity provide
us with a self--consistent way of modelling a possible variation in
Newton's constant, $G$.  In these theories the metric
tensor field of general relativity (GR) is no longer the only field mediating the
gravitational interaction; there is an additional scalar component.
Test particles in these theories still follow geodesics of the
metric but the manner in which the metric is generated from the
sources is altered by the
presence of the scalar field.  Such theories were first considered by
Jordan in 1949 \cite{Jor49} and later developed by Brans and Dicke in 1961
\cite{Bra61} for the case of a constant coupling parameter.  The
more general case of a dynamic coupling was considered by Bergmann \cite{Ber68},
Nordtvedt \cite{Nor70} and Wagoner \cite{Wag70} in the 1960s and
70s.  Solution--generating procedures for these theories were found by
Barrow and Mimoso \cite{Bar94}, Barrow and Parsons \cite{Bar97} and
Barrow \cite{Bar92}.

Using these theories we investigate the earliest well understood
physical process, primordial nucleosynthesis.  Previous studies on this subject have been carried out by a number of
authors.  In particular, the Brans--Dicke (BD) theory has been especially
well studied in this context by, for example, Casas, Garcia--Bellido and Quiros
\cite{Cas91}, \cite{Cas92} and Serna, Dominguez--Tenreiro and Yepes
\cite{Ser92}.  The more general class of ST theories has also been well
studied, most notably by Serna and Alimi \cite{Ser96}, Santiago, Kalligas and Wagoner
\cite{San97} and Damour and Pichon \cite{Dam99}.  More
  recently, a number of studies have been performed investigating the effect of a nonminimally
coupled quintessence field on primordial nucleosynthesis (see
e.g. \cite{Che01},\cite{Pet04}).  Although
these studies are very detailed they all make the simplifying assumption of a
constant $G$ during the radiation--dominated phase of the Universe's
history (with the exception of \cite{Ser92}, who
numerically investigate the effect of an early scalar--dominated phase
on the BD theory, and \cite{Dam99} who use the idea of a
``kick'' on the scalar field during electron--positron annihilation).  We relax this assumption and investigate the effects of entering the
radiation--dominated phase with a nonconstant $G$.  The constraints we impose
upon the variation of $G$ during primordial nucleosynthesis are then used to
constrain the parameters of the theory.  In carrying out this study we
consider the more general class of ST theories, paying
particular attention to the BD theory.

\section{Scalar--Tensor Gravity}

The ST theories of gravity are described by the action
\begin{multline}
S = \frac{1}{16\pi} \int \,d^4x
\sqrt{-g}(\phi R + \frac{\omega(\phi)}{\phi}g^{\mu\nu} \partial_\mu \phi
\partial_\nu \phi \\+16 \pi L_m).
\label{action}
\end{multline}
Here, $\phi$ is the scalar field, $R={R^{\mu}}_{\mu}$ is the Ricci
scalar and $L_m$ is the Lagrangian describing matter fields in the
space--time.  The action above can be extremized with respect to $g_{\mu\nu}$ to give the field equations
\begin{multline}
\label{fieldequations}
R^{\mu\nu}-\frac{1}{2} g^{\mu\nu} R + \frac{1}{\phi} (g^{\mu\rho}
g^{\nu\sigma} - g^{\mu\nu} g^{\rho\sigma}) \phi_{;\rho\sigma}\\ 
+ \frac{\omega(\phi)}{\phi^2} (g^{\mu\rho} g^{\nu\sigma} - \frac{1}{2}
g^{\mu\nu} g^{\rho\sigma}) \phi_{,\rho} \phi_{,\sigma}= -
\frac{8\pi}{\phi} T^{\mu\nu}
\end{multline}
and with respect to $\phi$ to give scalar--field propagation equation
\begin{equation}
\label{KleinGordon}
\square \phi = \frac{1}{2\omega(\phi) +3} (8 \pi T - \omega'(\phi)
g^{\mu\nu}\phi,_{\mu} \phi,_{\nu})
\end{equation}
where a prime denotes differentiation with respect to $\phi$,
$T^{\mu\nu}$ is the energy--momentum tensor of the matter fields, defined in the normal
way, and $T={T^{\mu}}_{\mu}$.  In the limit $\omega \rightarrow
\infty$ and $\omega'/\omega^3 \rightarrow 0$ the ST theories reduce to GR.

A useful property of these theories is that they can be conformally
transformed to a frame where the field equations take the same form as
in GR, with a nonminimally coupled, massless scalar field.  We refer to these
two frames as the Jordan (J) frame and the Einstein (E) frame,
respectively.  Under the conformal transformation
$g_{\mu\nu} = A^2(\phi) \bar{g}_{\mu\nu}$,
where $A^2(\phi)=1/\phi$, the Einstein --Hilbert action for GR containing a
scalar field, $\psi$, and other matter fields, $\Psi_m$,
\begin{multline}
S = \int\,d^4x \sqrt{-\bar{g}}  \left(\frac{1}{16\pi} \bar{R} + \frac{1}{2} \bar{g}^{\mu\nu}
\psi,_\mu \psi,_\nu\right)\\
+S_m[\Psi_m,A^2\bar{g}_{\mu\nu}].
\label{GR}
\end{multline}
is equivalent to the ST action (\ref{action}), if we make the definitions
$\Gamma \equiv \ln A(\phi)$, $\sqrt{4\pi}\alpha=\partial \Gamma/
\partial \psi$ and the identification $\alpha^{-2}=2\omega(\phi)+3$.
Here, symbols with bars refer to quantities in the E frame and symbols
without bars refer to quantities in the J frame.  

By extremizing the action (\ref{GR}) with respect to $\bar{g}_{\mu\nu}$ we get the E--frame field equations
\begin{equation*}
\bar{R}_{\mu\nu}=-8\pi(\bar{T}_{\mu\nu}-\frac{1}{2}\bar{g}_{\mu\nu}\bar{T}+\psi_{,\mu}\psi_{,\nu})
\end{equation*}
and by extremizing with respect to $\psi$ we get the E frame propagation equation
\begin{equation*}
\square\psi=-\sqrt{4\pi}\alpha\bar{T},
\end{equation*}
where we have defined the energy--momentum tensor in the E frame with
respect to $\bar{g}_{\mu \nu}$ so that $\bar{T}^{\mu\nu}=A^6T^{\mu\nu}$.

It is worth noting that whilst the J--frame energy--momentum tensor is always
covariantly conserved, $T^{\mu \nu}_{\quad ; \nu}=0$, its counterpart in the
E frame is not, $\bar{T}^{\mu \nu}_{\quad ; \nu}=\sqrt{4 \pi} \alpha \bar{T}
\psi ,^{\mu}$.  By insisting that in a physical frame the
energy--momentum of the matter fields should always be conserved we are forced to identify the J frame as the
physical one.  Any results derived in the E frame must be transformed
appropriately, if they are to be considered as physical.

\section{Cosmological Solutions}

\subsection{Brans--Dicke theory, $\omega=$constant}

Consider the homogeneous and isotropic
space--times described by the J--frame Friedmann--Robertson--Walker (FRW) line--element
\begin{equation}
\label{FRW}
ds^2 = dt^2 - a^2(t) \left( \frac{dr^2}{1-kr^2} + r^2 d\Omega^2
\right),
\end{equation}
where $k=0$, $\pm 1$ is the curvature parameter.

Substituting (\ref{FRW}) into (\ref{fieldequations}) and
(\ref{KleinGordon}) gives, for a perfect fluid, the J--frame Friedmann equations
\begin{equation}
\label{ten}
2\frac{\ddot{a}}{a}+\left(\frac{\dot{a}}{a}\right)^{2}+\frac{\omega }{2}\frac{\dot{\phi}^{2}}{\phi ^{2}}+2\frac{\dot{a}}{a}\frac{%
\dot{\phi}}{\phi }+\frac{\ddot{\phi}}{\phi }=-\frac{8\pi }{\phi }p-\frac{k}{%
a^{2}},
\end{equation}
\begin{equation}
\frac{8\pi }{3\phi }\rho =\left(\frac{\dot{a}}{a}\right)^{2}+\frac{\dot{a}}{a}\frac{\dot{\phi}}{\phi }-\frac{\omega }{6}%
\frac{\dot{\phi}^{2}}{\phi ^{2}}+\frac{k}{a^{2}},
\label{JFriedmann1and2}
\end{equation}
and
\begin{equation}
\label{JKG}
\frac{\ddot{\phi}}{\phi }=\frac{8\pi }{\phi }\frac{(\rho -3p)}{(2\omega +3)}%
-3\frac{\dot{a}}{a}\frac{\dot{\phi}}{\phi },
\end{equation}
where overdots denote
differentiation with respect to $t$.  From ${T^{\mu\nu}}_{;\nu}=0$ we
also obtain the perfect fluid conservation equation
\begin{equation*}
\dot{\rho}+3\frac{\dot{a}}{a}(\rho +p)=0.  \label{fluid}
\end{equation*}

\subsubsection{Radiation domination}

We consider first the case of a flat radiation--dominated universe.
Assuming the equation of state $p=\frac{1}{3} \rho$ and defining the
conformal time variable, $\eta$, by $a d\eta \equiv dt$, the equations
(\ref{ten}), (\ref{JFriedmann1and2}) and (\ref{JKG}) integrate to give
\cite{Gur72}
\begin{align}
\label{eta2}
\frac{a'(\eta)}{a(\eta)}&=\frac{\eta +\eta_2}{\eta^2+(2 \eta_2 +3
  \eta_1)\eta +\eta_2^2 + 3 \eta_1 \eta_2 -\frac{3}{2}\omega
  \eta_1^2}\\
\frac{\phi'(\eta)}{\phi(\eta)}&=\frac{3 \eta_1}{\eta^2+(2 \eta_2 +3
  \eta_1)\eta +\eta_2^2 + 3 \eta_1 \eta_2 -\frac{3}{2}\omega
  \eta_1^2}
\label{eta1}
\end{align}
where $\eta_1$ and $\eta_2$ are integration constants.

For $\omega > -3/2$ the solutions to these equations are
\begin{align}
a(\eta)&=a_1 (\eta +\eta_{+})^{\frac{1}{2}+
  \frac{1}{2\sqrt{1+\frac{2}{3} \omega}}} (\eta
  +\eta_{-})^{\frac{1}{2} -
  \frac{1}{2\sqrt{1+\frac{2}{3} \omega}}} \label{aw>}\\
\phi(\eta)&=\phi_1 (\eta +\eta_{+})^{-
  \frac{1}{2\sqrt{1+\frac{2}{3} \omega}}} (\eta +\eta_{-})^{+
  \frac{1}{2\sqrt{1+\frac{2}{3} \omega}}} \label{phiw>}
\end{align}
where $\eta_{\pm}=\eta_2+\frac{3}{2}\eta_1 \pm \frac{3}{2} \eta_1
\sqrt{1+\frac{2}{3}\omega}$, $a_1$ and $\phi_1$ are integration constants, and
$8 \pi \rho_{r0}/3 \phi_1 a_1^2=1$ (subscript 0 indicates a quantity
measured at the present day and we rescale so that $a_0=1$, throughout).
\begin{figure}
\epsfig{file=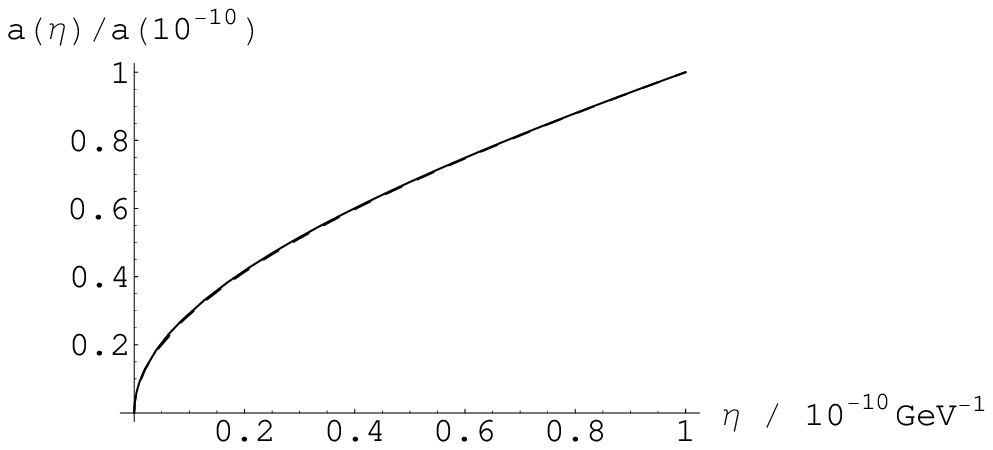,height=5.2cm}
\epsfig{file=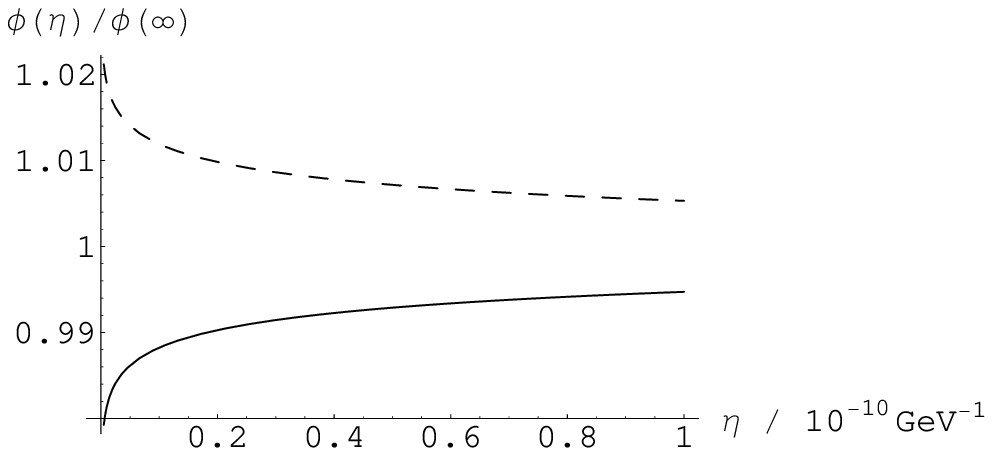,height=5.2cm}
\caption{{\protect {\textit{The evolution of $a$ and $\phi$ as functions of $\eta$
	for $\omega > -\frac{3}{2}$ with $\vert \eta_1 \vert=10^{-12} GeV^{-1}$,
	$\omega=40 000$, and $\eta_2$ set so that $a(0)=0$.  The solid
	line corresponds to $\eta_1 >0$ and the dashed line to  $\eta_1 <0$.}}}}
\label{raddom}
\end{figure}

For $\omega < -3/2$ we find
\begin{align}
a(\eta)&=a_1 \exp
\left(\frac{-\pi}{2\sqrt{\frac{2}{3}\vert\omega\vert-1}}\right) 
\sqrt{(\eta+\eta_{-})^2+\eta_{+}^2} \nonumber \\
&\qquad \qquad \; \times \exp \left(
  \frac{-1}{\sqrt{\frac{2}{3}\vert\omega\vert -1}} \tan^{-1}
  \frac{\eta+\eta_{-}}{\eta_{+}} \right), \label{aw<}\\
\phi(\eta)&=\phi_1 \exp\left(
\frac{\pi}{\sqrt{\frac{2}{3}\vert\omega\vert-1}}\right) \nonumber\\
&\qquad \qquad \; \times \exp \left(
  \frac{2}{\sqrt{\frac{2}{3}\vert\omega\vert -1}} \tan^{-1}
  \frac{\eta+\eta_{-}}{\eta_{+}} \right), \label{phiw<}
\end{align}
where $\eta_{+}= \frac{3}{2} \eta_1 \sqrt{
  \frac{2}{3}\vert\omega\vert-1}$, $\eta_{-}=\eta_2+\frac{3}{2} \eta_1$ and
  $8 \pi \rho_{r0}/3 \phi_1 a_1^2=1$.
\begin{figure}
\epsfig{file=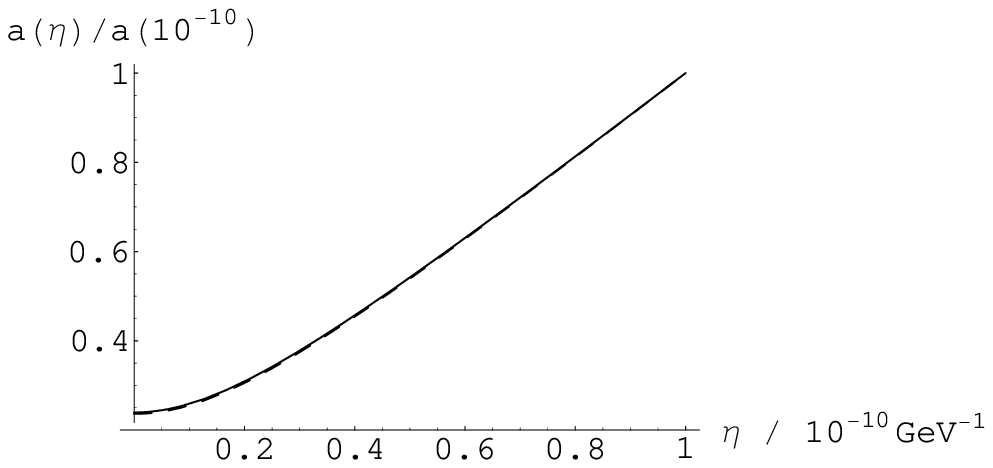,height=5.2cm}
\epsfig{file=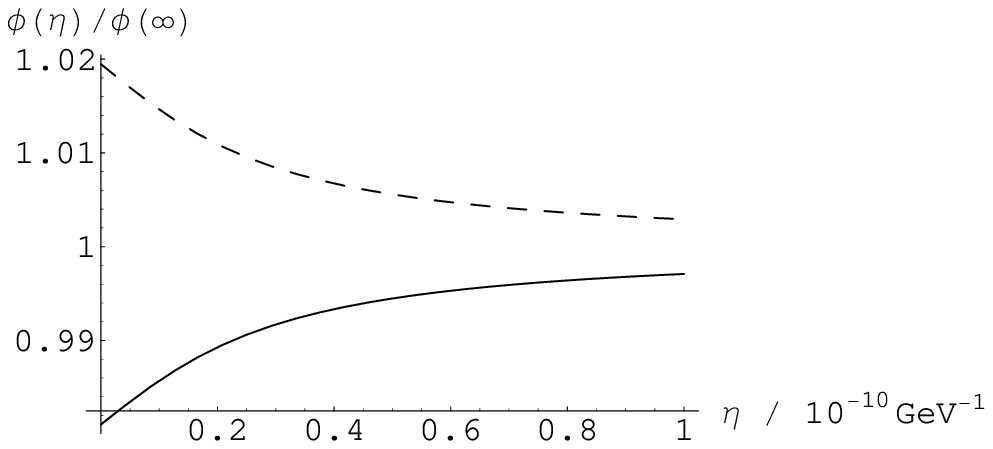,height=5.2cm}
\caption{\textit{The evolution of $a$ and $\phi$ as functions of $\eta$
	for $\omega < -\frac{3}{2}$ with $\vert \eta_1 \vert=10^{-13} GeV^{-1}$,
	$\omega=40 000$, and $\eta_2$ set so that $a'(0)=0$.  The solid
	line corresponds to $\eta_1 >0$ and the dashed line to  $\eta_1 <0$.}}
\label{raddom2}
\end{figure}

In these solutions $\eta_1$ determines the evolution of the scalar
field during radiation domination, and is a physically interesting
quantity; $\eta_2$ sets the origin of the conformal time coordinate.
The evolution of $a$ and $\phi$, for $\omega > - 3/2$ and $\omega < -3/2$ , is
shown in figures \ref{raddom} and \ref{raddom2}, respectively.  A
value of $\omega =40000$ is chosen, in agreement with recent solar
system observations \cite{Ber03}.

For $\omega >-3/2$ ($\omega <-3/2$) we see that the scale factor here
undergoes an initial period of rapid (slow)
expansion and at late times is attracted towards the solution $a(\eta)
\propto \eta$, or $a(t) \propto t^{\frac{1}{2}}$.  Similarly, $\phi$ can be seen to be
changing rapidly at early times and slowly at late times.  We
attribute these two different behaviours, at early and at late times, to
periods of free scalar--field domination and radiation domination,
respectively.  In fact, setting $\eta_1=0$ in (\ref{eta2}) and
(\ref{eta1}), we remove the scalar--dominated period and gain the
power--law exact solutions $a(\eta) \propto \eta$ and
$\phi=$ constant.  These are the ``Machian'' solutions usually considered in the
literature.  Allowing $\eta_1$ to be nonzero we will have a
nonconstant $\phi$, and hence $G$, during primordial
nucleosynthesis.  If $\rho_{r0}=0$ is chosen then these solutions become vacuum ones that are
driven by the $\phi$ field.  For $k=0$ these solutions do not have
GR counterparts. 

We see from (\ref{phiw<}), and figure \ref{raddom2},
that for $\omega<-3/2$ the initial singularity is avoided; for a more
detailed discussion of this effect see \cite{Bar04b}.

\subsubsection{Matter domination}

When considering a matter--dominated universe we could proceed as above and
determine a set of general solutions to (\ref{ten}), (\ref{JFriedmann1and2}) and (\ref{JKG})
that at early times are described by free scalar--field domination and at
late times by matter domination (see \cite{Gur72}), but for our
purposes this is unnecessary.  For a realistic universe we require a period of
radiation domination during which primordial nucleosynthesis can
occur.  If the scalar--field--dominated period of the Universe's history were to
impinge upon the usual matter--dominated period then we would effectively
lose the radiation--dominated era.  For this reason it is sufficient to
ignore the free scalar component of the general solution and consider
only the Machian component.  This is equivalent to imposing the
condition $\dot{\phi} a^3 \rightarrow 0$ as $a \rightarrow 0$.  With
this additional constraint the solutions to (\ref{ten}), (\ref{JFriedmann1and2}) and
(\ref{JKG}), for $k=0$ and $p=0$, are given by \cite{Nar68}
\begin{equation}
a(t) = a_* t^{\frac{2+2\omega}{4+3\omega}} \quad \text{and} \quad
\phi(t) =\phi_* t^{\frac{2}{4+3\omega}}
\label{matBD}
\end{equation}
where 
\begin{equation*}
\frac{8 \pi}{3 \phi_*} \frac{\rho_{m0}}{a_*^3} = \frac{2 (3+2
  \omega)}{3 (4+3 \omega)}  
\end{equation*}
and $a_*$ and $\phi_*$ are constants. These solutions can be seen to approach their GR counterparts as $\omega
\rightarrow \infty$.

\subsubsection{Vacuum domination}

Similarly, for a vacuum ($p=-\rho$) dominated period of expansion we can
impose the condition $\dot{\phi} a^3 \rightarrow 0$ as $a
\rightarrow 0$ to get the power law exact solutions \cite{Nar68} 
\begin{equation}
\label{vacBD}
a(t) =a_{\dagger} t^{\frac{1}{2}+\omega} \quad \text{and} \quad \phi(t) =\phi_{\dagger} t^2
\end{equation}
where 
\begin{equation*}
\frac{8 \pi}{3 \phi_{\dagger}} \rho_{\Lambda 0} = \frac{1}{12}
(3+2\omega) (5+6\omega).  
\end{equation*}
and $a_{\dagger}$ and $\phi_{\dagger}$ are constants.  Note that this vacuum stress does not
produce a de Sitter metric as in GR.

\subsection{Dynamically--coupled theories, $\omega=\omega(\phi)$}

In order to evaluate more general scalar--tensor theories it is
convenient to work in the E frame.  In this frame, substituting the FRW
line--element into the field equations gives
\begin{equation}
\label{EFriedmann1and2}
\left( \frac{\dot{\bar{a}}}{\bar{a}}\right)^2 =
\frac{8\pi}{3}(\bar{\rho}+\frac{1}{2}\dot{\psi}^2)-\frac{k}{\bar{a}^2},
\end{equation}
\begin{equation}
\label{25}
\frac{\ddot{\bar{a}}}{\bar{a}} =
-\frac{4\pi}{3}(\bar{\rho}+3\bar{p}+2\dot{\psi}^2)
\end{equation}
and
\begin{equation}
\ddot{\psi}+3\frac{\dot{\bar{a}}}{\bar{a}}\dot{\psi} = -\sqrt{4\pi}\alpha(\bar{\rho}-3\bar{p})
\label{EFriedmann3}
\end{equation}
where over--dots here denote differentiation with respect to $\bar{t}$
and $\bar{a}(\bar{t})=A(\phi) a(t)$ is the scale--factor in the E
frame.  Defining $N= \ln (\bar{a}/\bar{a}_0)$ Damour and Nordtvedt \cite{Dam93} write
(\ref{EFriedmann1and2}), (\ref{25}) and (\ref{EFriedmann3}) as
\begin{equation}
\label{master}
\frac{2(1-\epsilon)}{(3-4\pi\psi'^2)}\psi''+(2-\gamma-\frac{4}{3}\epsilon)\psi'=-(4-3\gamma)\frac{\alpha}{\sqrt{4\pi}},
\end{equation}
where $\epsilon=3k/8\pi\bar{\rho}\bar{a}^2$, $\bar{p}=(\gamma-1) \bar{\rho}$ and
$'$ denotes differentiation with respect to $N$.

Now we consider a coupling parameter of the form
\begin{equation*}
\Gamma(\psi) =\Gamma(\psi_{\infty})+ \alpha_0 (\psi(\bar{t})-\psi_{\infty})+  \frac{1}{2}\beta(\psi(\bar{t})-\psi_{\infty})^2
\end{equation*}
so that
\begin{equation}
\label{alpha}
\alpha(\bar{t})=\alpha_0 +\frac{\beta}{\sqrt{4\pi}}(\psi(\bar{t})-\psi_{\infty}).
\end{equation}
This is of the same form as chosen by Santiago, Kalligas and Wagoner
\cite{San97} and Damour and Pichon \cite{Dam99}.  Santiago,
Kalligas and Wagoner arrive at this form of $\alpha$ by assuming
the evolution of the Universe has been close to the GR solutions
throughout the period from primordial nucleosynthesis to the present.  They
therefore consider themselves justified in
performing a Taylor expansion about the asymptotic value of $\psi$ and
discarding terms of second order or higher.  Damour and Pichon consider
$\Gamma$ to be a potential down which $\psi$ runs.  They assume that a particle near
a minimum of a potential experiences a generically parabolic form for
that potential. This leads them to consider a quadratic form for $\Gamma$
which gives, on differentiation, $\alpha$ as above.  These two lines
of reasoning are, of course, equivalent as the parabolic form of a
particle near its minimum of potential can be found using a Taylor
series.  Theories with this form of $\alpha$ belong to the attractor
class which approach GR at late times, if we impose the additional
condition $\alpha_0=0$.

\subsubsection{Radiation domination}

For the case of a radiation--dominated flat universe, the general
solutions of (\ref{EFriedmann1and2}), (\ref{25}) and
(\ref{EFriedmann3}) with the
choice (\ref{alpha}), are, for $\omega > -3/2$:
\begin{align*}
\bar{a}^2(\eta) &= \frac{8 \pi \bar{\rho}_{r0}}{3} (\eta-\eta_2)
(\eta-\eta_2+\vert \eta_1 \vert)\\
\psi-\psi_1 &= \sqrt{\frac{3}{16 \pi}} \frac{\eta_1}{\vert \eta_1 \vert} \ln
  \left( \frac{\eta -\eta_2}{\eta -\eta_2+\vert
    \eta_1 \vert} \right),
\end{align*}
and, for $\omega < -3/2$,
\begin{align*}
\bar{a}^2(\eta) &= \frac{2 \pi \bar{\rho}_{r0}}{3} \vert \eta_1 \vert^2 +
\frac{8 \pi \bar{\rho}_{r0}}{3} (\eta-\eta_2)^2\\
\psi-\psi_1 &= -\sqrt{\frac{3 \pi}{16}} \frac{\eta_1}{\vert \eta_1 \vert} +
\sqrt{\frac{3}{4 \pi}} \frac{\eta_1}{\vert \eta_1 \vert} \tan^{-1}
\left( \frac{2(\eta -\eta_2)}{\vert \eta_1 \vert} \right).
\end{align*}
Here, $\eta_1$ determines the evolution of the scalar
field, $\psi_1$ is the value it approaches asymptotically and $\eta_2$ sets the origin of the conformal time
coordinate defined by $\bar{a}(\eta) d \eta=d \bar{t}$.
For $\omega > -3/2$, $\eta_1 \in \mathcal{R}$ such that $\psi \in
\mathcal{R}$ whilst for $\omega < -3/2$, $\eta_1 \in \mathcal{I}$ such
that $\psi \in \mathcal{I}$ and $\omega \in \mathcal{R}$.  If the
idea of an imaginary scalar field seems strange the reader is
reminded that the E frame is considered to be unphysical: all
observable quantities measured in the physical J frame remain real at all times.

These solutions in the J frame are
\begin{equation}
\phi(\eta) = \exp \left( -\beta (\psi(\eta)-\psi_{\infty})^2 \right)
\label{STrad}
\end{equation}
and
\begin{equation}
a^2(\eta) = \frac{\bar{a}^2(\eta)}{\phi}. \label{STrad2}
\end{equation}
They exhibit the same features as their BD
counterparts: at early times there is a period of free--scalar--field
domination, and at late times they approach $a \propto t^{\frac{1}{2}}$ and $\phi
=$constant.

\subsubsection{Matter domination}

Solutions during the matter--dominated era are difficult to find because
the energy--momentum tensor for the matter field is not
conserved.  However, it is possible to obtain an evolution
equation for $\psi$.  Using (\ref{master}), we get 
\begin{equation}
\label{matter}
\frac{2(1-\epsilon)}{(3-4\pi\psi'^2)}\psi''+(1-\frac{4}{3}\epsilon)\psi'=
-\frac{\beta}{4\pi}(\psi-\psi_{\infty}).
\end{equation}
For a flat universe $\epsilon=0$ and we can solve
(\ref{matter}) by making the simplifying assumption
$\psi'^2\ll 3$.  This gives the solution
\begin{multline}
\label{solution}
\psi(N)-\psi_{\infty}=
A e^{-\frac{3}{4}(1+\sqrt{1-\frac{2\beta}{3\pi}})N}\\
+B e^{-\frac{3}{4}(1-\sqrt{1-\frac{2\beta}{3\pi}})N}
\end{multline}
where $A$ and $B$ are constants of integration.

\subsubsection{Vacuum domination}

Similarly, for the case of a vacuum--dominated universe the evolution
of $\psi$ can be approximated using (\ref{master}) to obtain
\begin{multline}
\label{solution2}
\psi(N)-\psi_{\infty}=
Ce^{-\frac{3}{2}(1+\sqrt{1-\frac{2\beta}{3\pi}})N}\\
+De^{-\frac{3}{2}(1-\sqrt{1-\frac{2\beta}{3\pi}})N}
\end{multline}
where C and D are constants of integration.

\section{Primordial Nucleosynthesis}

\subsection{Modelling the form of $G(t)$}

If primordial nucleosynthesis were to occur during the
scalar--dominated period then the very different expansion rate would have disastrous
consequences for the light--element abundances (see e.g. \cite{Ser92}).
Therefore we limit our study to times at which the scale factor can be
approximated by a form that is close to $a(\eta) \propto \eta$, and so
primordial nucleosynthesis can safely be described as occurring during radiation
domination.  Performing a power--series expansion of the solutions
(\ref{aw>}), (\ref{phiw>}), (\ref{aw<}) and (\ref{phiw<}) in
$\eta_1/(\eta+\eta_2)$ we find
\begin{align}
a(\eta) &= a_1 (\eta+\eta_2 +3 \eta_1) +O(\eta_1^2) \label{a1}\\
\phi(\eta) &= \phi_1 \left( 1-\frac{3 \eta_1}{\eta+\eta_2} \right)
+O(\eta_1^2) \label{a2}
\end{align}
for both $\omega >-3/2$ and $\omega <-3/2$.  We can then set the
origin of the $\eta$ coordinate such that $a(0)=0$ with the choice
$\eta_2=-3 \eta_1$.  The solutions (\ref{a1}) and ({\ref{a2}) then
  become, in terms of the time coordinate $t$,
\begin{align*}
a(t) &= a_1 t^{\frac{1}{2}} +O(\eta_1^2)\\
\phi(t) &= \phi_1 \left(1+\frac{a_2}{a(t)} \right) +O(\eta_1^2),
\end{align*}
where $a_2=-3 \eta_1 a_1$ and the origin of $t$ has been chosen to coincide with the origin of
$\eta$.

Similarly, expanding the solutions (\ref{STrad}) and (\ref{STrad2}) in
$\eta_1/(\eta+\eta_2)$ we find that, for both $\omega >-3/2$ and $\omega <-3/2$,
\begin{align*}
a(t) &= a_1 t^{\frac{1}{2}} + O(\eta_1^2)\\
\phi(t) &= \phi_1 \left( 1+\frac{a_2}{a(t)} \right) +O(\eta_1^2)
\end{align*}
where 
\begin{align*}
a_1 &= \sqrt{8 \pi \bar{\rho}_{r0}/3} \exp
(\beta(\psi_1-\psi_{\infty})^2/2)\\
a_2 &=  \sqrt{2 \bar{\rho}_{r0}}\beta \eta_1
(\psi_1-\psi_{\infty})  \exp
(\beta(\psi_1-\psi_{\infty})^2/2)\\
\phi_1 &= \exp (-\beta
(\psi_1-\psi_{\infty})^2)
\end{align*}
and $\eta_2$ has been set so that
$a(0)=0$.

In the limit $\dot{\phi} \rightarrow 0$ we can see from equations
(\ref{JFriedmann1and2}) that the standard GR Friedmann equations are
recovered, with a different value of Newton's constant given by
\begin{equation}
\label{Gt}
\begin{split}
G(t) = \frac{1}{\phi(t)} = G_1 \frac{a(t)}{a(t)+a_2}
\end{split}
\end{equation}
where $G_1=1/\phi_1$.  We conclude that the solutions found above correspond to a
situation that can be described using the GR Friedmann equations with a
different, and adiabatically changing, value of $G$.  This is just the
situation considered in recent work by Bambi, Giannotti and Villante
\cite{Bam05}, but we now have an explicit form for the evolution of $G(t)$
derived from ST gravity theory.

\begin{figure*}
\subfigure[2$\sigma$ bounds for $a_2 > 0$]{\epsfig{figure=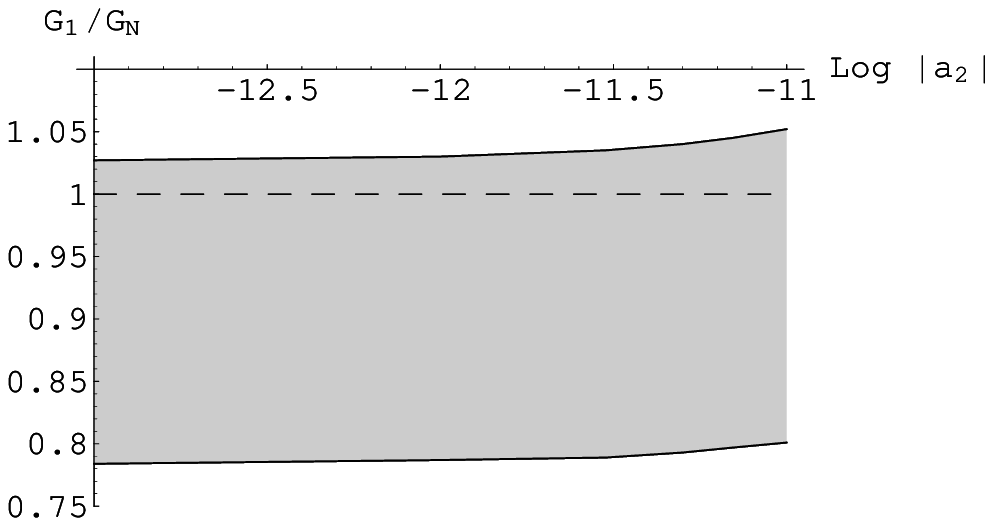, height=5.2cm}}
\subfigure[2$\sigma$ bounds for $a_2 < 0$]{\epsfig{figure=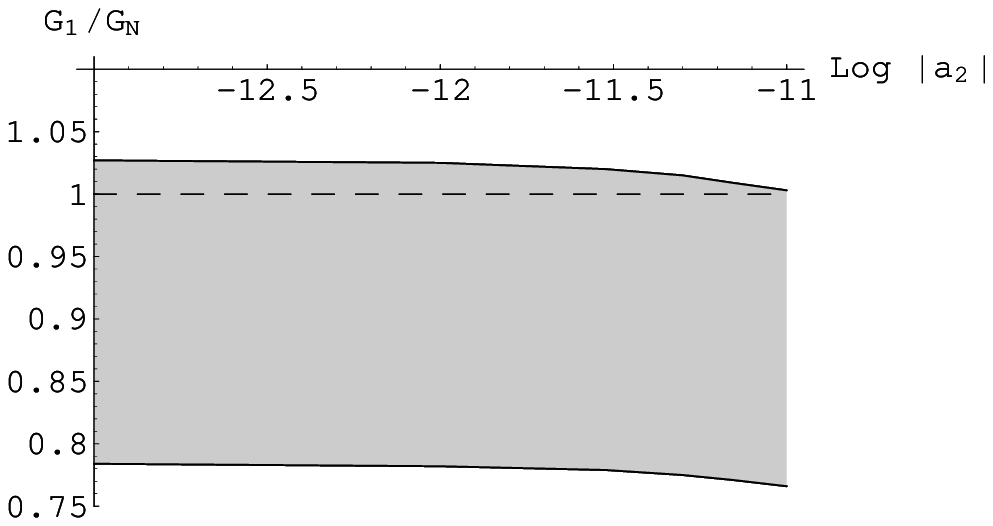, height=5.2cm}}
\caption{\textit{The upper and lower bounds on $G_1/G_N$ as a
    function of $a_2$ are shown by the solid lines.  The shaded regions correspond to the allowed
    parameter space and the dashed lines show $G_1=G_N$.}}
\label{scherrerplot}
\end{figure*}

\subsection{The effect of a non--constant $G$}

The time at which weak interactions freeze out in the early Universe
is determined by equality between
the rate of the relevant weak interactions and the Hubble rate.  When the
weak interaction rate is the greater then the ratio of neutrons
to protons tracks its equilibrium value, $n/p = \exp(-(m_n-m_p)/T)$
where $m_n$ and $m_p$ are the neutron and proton masses.
If, however, the Hubble rate is greater than the weak--interaction rate
then the ratio of neutrons to
protons is effectively ``frozen--in'', and $\beta$ decay is the only
weak process that still operates with any efficiency.  This will be the
case until the onset of deuterium formation, at which time the
neutrons become bound and $\beta$ decay ceases.  The onset of
deuterium formation is primarily determined by the photon to
baryon ratio, $\eta_{\gamma}$, which inhibits the formation of
deuterium nuclei until the critical temperature for photodissociation is past.  As the vast
majority of neutrons end up in $^4$He the primordial abundance of this
element is influenced most significantly by the number of neutrons at
the onset of deuterium formation, which is most sensitive to the
temperature of weak--interaction freeze--out, and hence the Hubble
rate, and so $G$, at this time.  Conversely, the primordial abundances of
the other light elements are most sensitive to the temperature of
deuterium formation, and hence $\eta_{\gamma}$, when
nuclear reactions occur and the light elements form.  (See
\cite{Bam05} for a more detailed discussion of these points).

Using the simple forms of $a(t)$ and $G(t)$ derived above
we use a modified version of the Kawano code \cite{Kaw92} to
investigate the effect of this variation of $G$ on primordial nucleosynthesis directly.  We
use the deuterium abundance estimated by Kirkman et al. \cite{Kir03}
\begin{equation}
\label{Dbound}
\log \left( \frac{D}{H} \right) = -4.556 \pm 0.064 \quad \text{to} \;
1\sigma,
\end{equation}
and the $^4$He abundance estimated by Barger et al. \cite{Bar03}
\begin{equation}
\label{Hebound}
Y_P=0.238 \pm 0.005 \quad \text{to} \; 1\sigma,
\end{equation}
to create a parameter space in ($\eta_{\gamma}$, $G_1$, $a_2$).  The three-dimensional 95$\%$ $\chi^2$ confidence region is projected into
the $G_1$, $a_2$ plane to give figure \ref{scherrerplot}, where three species of light neutrinos and a neutron
mean lifetime of $\tau=885.7$ seconds have been assumed.

The apparent disfavoring of the value of $G_1$ much greater than $G_N$
in figure \ref{scherrerplot} is due to the observational limits we
have adopted for $Y_P$ (equation (\ref{Hebound})), which
suggest a helium abundance that is already uncomfortably low compared to the 
theoretical prediction for $Y_P$ derived from the baryon density corresponding 
to equation \ref{Dbound}.  We expect these constraints to be
updated by future observations (see e.g. \cite{Tro04}); this
will require a corresponding update of any work, such as this,
that seeks to use these constraints to impose limits on physical
processes occuring during primordial nucleosynthesis.

These limits on the parameters $G_1$ and $a_2$ can be used to
construct plots showing the explicit evolution of $G(t)$ for various
limiting combinations of the two parameters, this is done in figure
\ref{G}.  It is interesting to note that in both of these plots the
lines corresponding to different values of $a_2$ all appear to cross at approximately the same point, $\log a \sim
-9.4$.  This confirms our earlier discussion, and the results of \cite{Bam05}, that
the $^4$He abundance is mostly only sensitive to the value of $G$ at the
time when the weak interactions freeze out.  In reality, this
freeze--out happens over a finite time interval, but from figure
\ref{G} we see that it is a good approximation to consider it happening
instantaneously - where the lines cross.  To a
reasonable accuracy one could then take $G$ throughout primordial nucleosynthesis
to be its value during the freeze--out process.  
\begin{figure*}
\subfigure[$a_2 > 0$]{\epsfig{file=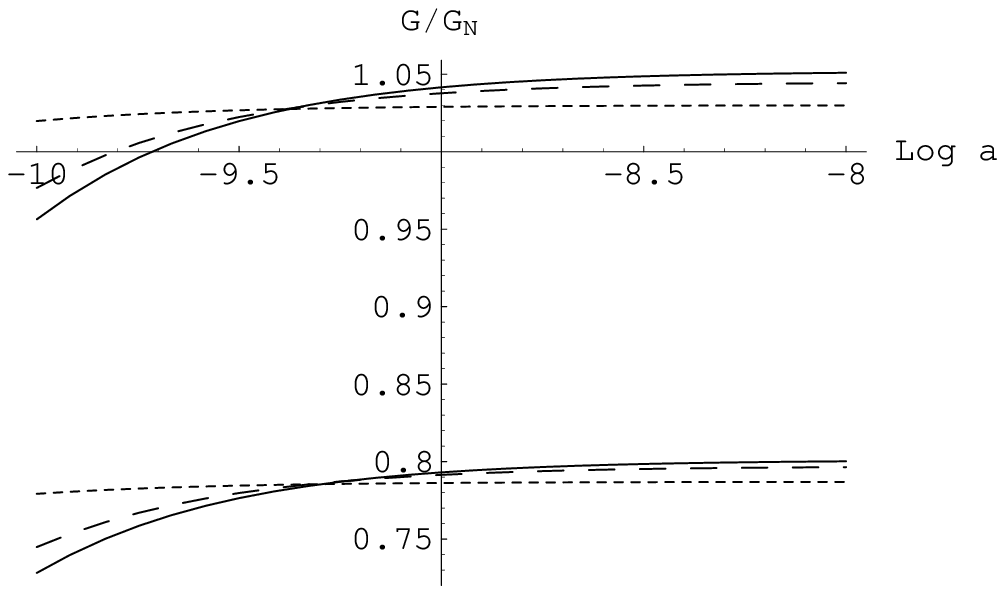,height=5.2cm}}
\subfigure[$a_2 < 0$]{\epsfig{file=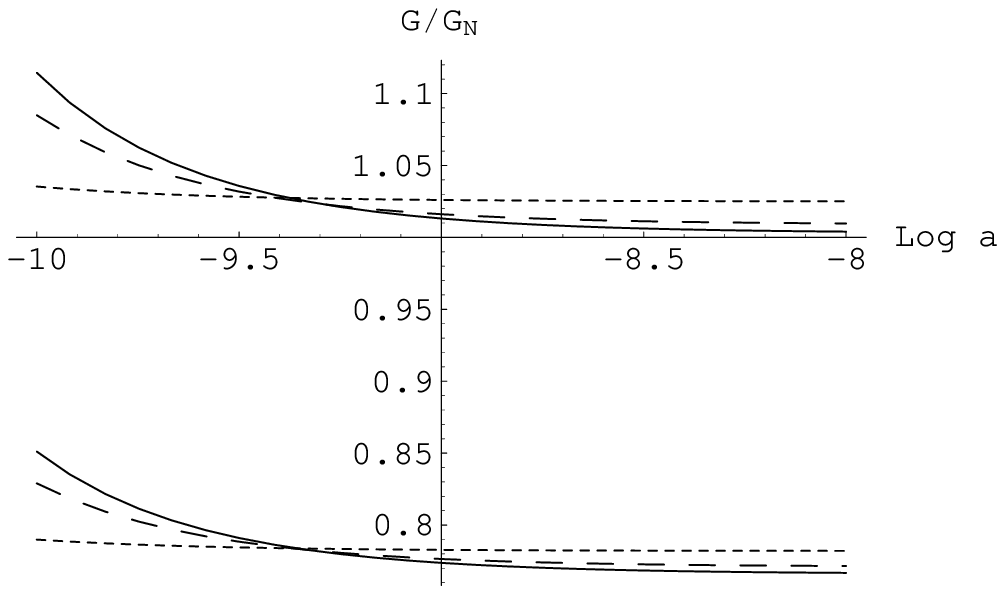,height=5.2cm}}
\caption{\textit{These plots show the explicit evolution of G for some
limiting values of $G_1$, as taken from figure \ref{scherrerplot}, for
various values of $a_2$.  Solid lines correspond to $\vert a_2 \vert =
10^{-11}$, dashed lines correspond to $\vert a_2 \vert =
10^{-11.155}$ and dotted lines correspond to $\vert a_2 \vert =
10^{-12}$.}}
\label{G}
\end{figure*}

\section{Constraining the theory}

Using the results in the previous section it is possible to constrain
the underlying ST gravitational theory.  This is done separately for the
BD theory and for the more general ST theories.
For each theory we consider the case of a universe containing matter
and radiation only and then a universe containing matter, radiation and a
non--zero vacuum energy, with equation of state $p=-\rho$.

\subsection{Brans--Dicke theory}

\subsubsection{Universe containing matter and radiation}

Using equations (\ref{matBD}) we can write the
ratio of $G$ at matter--radiation equality, $G_{eq}$, to its
present--day laboratory value, $G_N$, as
\begin{multline}
\frac{G_{eq}}{G_N}= \frac{(2 \omega+3)}{(2
  \omega+4)}\frac{\phi_N}{\phi_{eq}} = \frac{(2 \omega+3)}{(2
  \omega+4)}\left( \frac{a_0}{a_{eq}}\right)^{1/(\omega+1)}\\
\equiv  \frac{(2 \omega+3)}{(2 \omega+4)}(1+z_{eq})^{1/(\omega+1)}
\label{speedup}
\end{multline}
where we have used the expression for $G$ in the weak--field limit \cite{Wil93}
\begin{equation}
G(\phi)=\frac{(2\omega+4)}{(2\omega+3)}\frac{1}{\phi},
\label{JG}
\end{equation}
for $G_N$, whilst using the expression $G(\phi)=1/\phi$ for $G_{eq}$, as in
(\ref{Gt}).  In reality, the laboratory value of $G$ today is
not equal to that in the background cosmology \cite{Cli05}, although we
take it to be so here for simplicity.

We now proceed by calculating $1+z_{eq}$ in the
Brans--Dicke cosmology, following Liddle, Mazumdar and Barrow
\cite{Lid98}.  As $T^{\mu \nu}_{\quad ; \nu}=0$ in the J frame we have that $\rho_r\propto
a^{-4}$ and $\rho_m\propto a^{-3}$.
For a universe containing matter and radiation only this gives
\begin{equation*}
1+z_{eq}=\frac{a_0}{a_{eq}}=\frac{\rho_{m0}}{\rho_{r0}}\frac{\rho_{req}}{\rho_{meq}}
=\frac{\rho_{m0}}{\rho_{r0}}
\end{equation*}
as well as the usual relation $T \propto a^{-1}$, when entropy
increase is neglected.

Photons and neutrinos both contribute to the value of $\rho_{r0}$, the
present--day energy--density of radiation.  From $T_{\gamma 0}=2.728 \pm
0.004K$ \cite{Fix96}, we get $\rho_{\gamma 0}=4.66\times 10^{-34}
g \; cm^{-3}$ and using the well known result $T_{\nu 0}=
(4/11)^{1/3}T_{\gamma 0}$, and assuming three families of light neutrinos,
we also have $\rho_{\nu 0}=0.68\rho_{\gamma 0}$.  This gives the total
present--day radiation density as
$\rho_{r0}=7.84 \times 10^{-34}g \; cm^{-3}$.  Now, recalling our
assumption of spatial flatness and (\ref{JFriedmann1and2}), we can write
\begin{equation*}
\rho_{tot0}=\rho_{m0}+\rho_{r0}=\frac{3H_0^2}{8\pi G_N}\frac{(4+3\omega)(4+2\omega)}{6(1+\omega)^2}.
\end{equation*}
For $G_N=6.673\times10^{-11} \;  Nm^2kg^{-2}$, $H_0=100h \;
km s^{-1} Mpc^{-1}$, and the value of $\rho_{r0}$ above, we
have 
\begin{align*}
1+z_{eq} &=2.39 \times 10^4 h^2
\frac{(4+3\omega)(4+2\omega)}{6(1+\omega)^2} -1\\
&= 2.39 \times 10^4 h^2 \left( 1+ \frac{4}{3 \omega} \right) +O(\omega^{-2}).
\end{align*}
This correction to $1+z_{eq}$ has direct observational consequences in the
power spectrum of cosmic microwave background (CMB) perturbations.  After $1+z_{eq}$ the
subhorizon scale perturbations, that were previously effectively
frozen, are allowed to grow.  Changing the value of $1+z_{eq}$
therefore causes a shift in the power--spectrum peaks, which is
potentially observable (see \cite{Lid98}, \cite{Che99}, \cite{Nag02}
\cite{Zah03}, \cite{Nag04} and \cite{Lid04} for a more detailed discussion of the effect of a varying
$G$ on CMB formation).  For our purposes this modified expression
for $1+z_{eq}$ can then be substituted into (\ref{speedup}) to give an equation
in terms of $\omega$, $G_{eq}$ and $h$.  Assuming a value of $h=0.7$
and that $G_{eq}=G_1$, as appears a very good approximation for the
models above, we can use our bounds on $G_1$ in terms of $a_2$ to
create an allowed parameter space in the ($\omega$,$a_2$) plane.  This
is shown in figure \ref{BDresult}.

We remind the reader that the a large $\omega$ corresponds to a slowly
varying Machian component of $\phi(1+z)$, as can be seen directly from
equations (\ref{matBD}), (\ref{vacBD}) and the late--time limits of (\ref{aw>}),
(\ref{phiw>}), (\ref{aw<}) and (\ref{phiw<}).  The parameter $a_2$ determines the
evolution of the free component of $\phi$, as can be seen from the
early time limits of (\ref{aw>}), (\ref{phiw>}), (\ref{aw<}) and (\ref{phiw<}).  In
the limits $\omega \rightarrow \infty$ and $a_2 \rightarrow 0$ the
Machian and free components of $\phi(1+z)$ both become constant,
respectively, resulting in a constant $G(1+z)$.  The constraints imposed
upon $\omega$ and $a_2$ in figure \ref{BDresult} therefore correspond
to constraints upon the evolution of $G(1+z)$ in this theory, valid both during priomordial
nucleosynthesis and at other cosmological epochs.

\subsubsection{Universe containing matter, radiation and a nonzero
  vacuum energy}

A more realistic constraint would involve taking into account a
late--time period of vacuum domination; so as well as $\rho_r\propto
a^{-4}$ and $\rho_m\propto a^{-3}$ we now also have $\rho_{\Lambda}
=\text{constant}$ in the J frame.  Hence,
\begin{align*}
\frac{G_{eq1}}{G_N} &=\frac{(2\omega+3)}{(2\omega+4)}\frac{\phi_0}{\phi_{eq1}} 
= \frac{(2\omega+3)}{(2\omega+4)} \frac{\phi_{eq2}}{\phi_{eq1}}
    \frac{\phi_{0}}{\phi_{eq2}}\\
&=\frac{(2\omega+3)}{(2\omega+4)}  \left(
    1+z_{eq1}\right)^{\frac{1}{(1+\omega)}}
    \left(1+z_{eq2} \right)^{-\frac{\omega}{(1+\omega)(1+2\omega)}}
\end{align*}  
where we have defined the redshift of matter--radiation equality, $z_{eq1}$, and
the redshift of matter--vacuum equality, $z_{eq2}$, as
\begin{equation*}
1+z_{eq1}=\frac{\rho_{m0}}{\rho_{r0}}
\quad \text{and} \quad
1+z_{eq2}=\left(\frac{\rho_{\Lambda0}}{\rho_{m0}}
\right)^{\frac{1}{3}}.  
\end{equation*}
As above, we still have $\rho_{r0}=7.84 \times 10^{-34}g \; cm^{-3}$ but
now our assumption of spatial flatness gives
\begin{equation*}
\rho_{tot0}=\rho_{\Lambda0}+\rho_{m0}+\rho_{r0}=\frac{3H_0^2}{8\pi
  G_N}\frac{(4+3\omega)(4+2\omega)}{6(1+\omega)^2} .
\end{equation*}
or, using $G_N=6.673\times10^{-11} \;  Nm^2kg^{-2}$ and $H_0=100h \;
km s^{-1} Mpc^{-1}$,
\begin{equation*}
1+z_{eq1}=\frac{\left( 2.39 \times 10^4 h^2
\frac{(4+3\omega)(4+2\omega)}{6(1+\omega)^2} -1 \right)}{\left( 1+\frac{\rho_{\Lambda 0}}{\rho_{m0}} \right)}.
\end{equation*} 
Taking the value $\rho_{\Lambda0}/\rho_{m0}=2.7$, consistent with
WMAP observations \cite{Ben03}, we get the results shown in \ref{BDresult}.
\begin{figure}
\epsfig{file=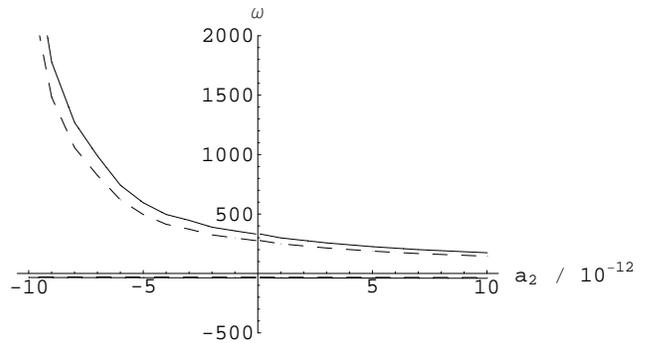,height=5.2cm}
\caption{\textit{The allowed parameter space in this plot is the
    region above the higher
    solid line and below the lower solid line, with $h=0.7$ to 2$\sigma$
    for a universe containing matter and radiation only.  The dashed
    lines show the corresponding result for a universe with a
    non--zero vacuum energy.}}
\label{BDresult}
\end{figure}

\subsection{Dynamically--coupled theories}

\subsubsection{Constraints on $\omega(\phi)$ at matter--radiation equality}

Recalling that $A^2=1/\phi$ and $\alpha^{-2}=2\omega+3$ allows us to re--write (\ref{JG}) as
\begin{equation*}
G=A^2(1+\alpha^2),
\end{equation*}
the ratio $G_{eq}/G_N$ can then be expressed in terms of $A$ and $\alpha$ as
\begin{equation*}
\frac{G_{eq}}{G_N}=\frac{A_{eq}}{A_0\sqrt{1+\alpha_0^2}}.
\end{equation*}
Assuming that $\ln A_0$ and $\ln (1+\alpha_0)$ are negligible
compared to $\ln A_{eq}$ (i.e. the Universe is close to GR
today, \cite{Ber03}) we can write
\begin{align*}
\ln \left( \frac{G_{eq}}{G_N} \right) \simeq \ln A_{eq} &=
\frac{1}{2}\beta (\psi_{eq}-\psi_{\infty})^2 \\ &=\frac{2 \pi}{(2
  \omega_{eq}+3)} \frac{1}{\beta}.
\end{align*}
This allows us to constrain $\omega_{eq}$ in terms of $\beta$ and
$a_2$, as shown in figure \ref{STresult1}.  Constraints imposed upon
$\omega$ and $a_2$ have the same consequences for the evolution of $G$
  as previously discussed.  The parameter $\beta$ controls the
  evolution of $\omega$, as can be seen from (\ref{alpha}).  In the
  limit $\beta \rightarrow 0$ it can be seen that $\omega$ becomes constant
  and this class of ST theories becomes indistinguishable from the BD
  theory.  Constraints upon $\beta$ therefore correspond to
  constraints on the allowed variation of $\omega$ and hence $G$.

The parameter $\beta$ is
taken to be small here so that the ``kick'' on the scalar field during
electron--positron annihilation can be neglected.  These effects
have been explored by Damour and Pichon in \cite{Dam99}, for the case
$a_2=0$, and are expected to have the same result in this more general
scenario; we will not repeat their analysis of this effect here.
\begin{figure}
\epsfig{file=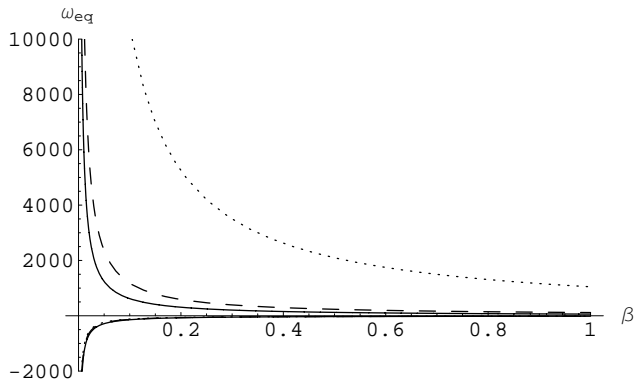,height=5.2cm}
\caption{\textit{The allowed parameter space in this graph is the region above the
    line in the region $\omega>0$ and below the line in the region
    $\omega<0$, to 2$\sigma$.  The solid lines correspond to
    $a_2=10^{-11}$ the dashed lines to $a_2=10^{-13}$ and the dotted
    lines to $a_2=-10^{-11}$.  $\omega_{eq}$ is the value of
    $\omega$ at matter--radiation equality and $\beta$ is a parameter
    of the theory, defined in (\ref{alpha}).}}
\label{STresult1}
\end{figure}

\subsubsection{Constraints on $\omega_0$ for a universe containing
    matter and radiation}

The scalar field can be evolved forward in time from the time of
matter--radiation equality to the present, using equation
(\ref{solution}).  The two arbitrary constants in this expression can be
fixed using our limiting values of $G_{eq}$ derived above and by assuming that
the evolution of the scalar field has effectively ceased by this time,
i.e. $\psi'_{eq}=0$, as is the case for the models considered here.

In order to gain a quantitative limit on $\psi_1-\psi_{\infty}$, and
hence on $\omega_0$, it is necessary to calculate $N_{eq}$.  From the
definition of $N$, we can write
\begin{equation*}
\begin{split}
N_{eq}&=-\ln (1+z_{eq})-\ln A_{eq} + \ln A_0\\
&\simeq -\ln (1+z_{eq})-\frac{1}{2}\beta(\psi_{eq}-\psi_{\infty})^2
\end{split}
\end{equation*}
where $z_{eq}$ is the red--shift at $t_{eq}$ and we have
assumed, as before, the term $\ln A_0$ to be negligible.

It now remains to determine 
$1+z_{eq}=\rho_{m0}/\rho_{r0}$ for the case $\omega=\omega(\phi)$.  If
we now assume $\omega_0$ to be moderately large, and recall
our assumption of spatial flatness, we can write
\begin{equation*}
\rho_{tot0}=\rho_{m0}+\rho_{r0} = \frac{3H_0^2}{8\pi G_N}.
\end{equation*}
For $\rho_{r0}=7.84 \times 10^{-34}g \; cm^{-3}$ this gives
$\rho_{m0} \simeq 1.87 \times 10^{-29} \; h^2 g cm^{-3}$.  We
have $1+z_{eq} \simeq 2.4 \times 10^4 \; h^2$ and so finally obtain, for $h=0.7$,
\begin{equation*}
N_{eq} \simeq -9.37-\frac{1}{2}\beta(\psi_{eq}-\psi_{\infty})^2.
\end{equation*}

Now, evolving $\psi-\psi_{\infty}$ from $N_{eq}$ to $N_0=0$ and using
\begin{equation}
\label{w(t)}
\omega_0=\frac{2\pi}{\beta^2(\psi_0-\psi_{\infty})^2}-\frac{3}{2}
\end{equation}
we obtain the results shown in figure \ref{STresult2}.
\begin{figure}
\epsfig{file=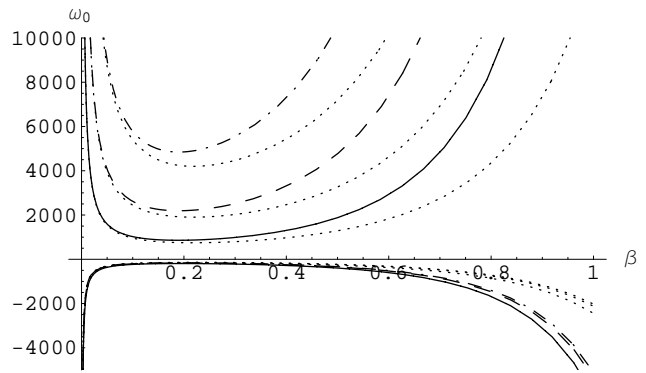,height=5.2cm}
\caption{\textit{The allowed parameter space in this graph is the region above the
    line in the region $\omega>0$ and below the line in the region
    $\omega<0$, to 2$\sigma$ for a universe containing matter and radiation only.  The solid lines correspond to
    $a_2=10^{-11}$, the dashed lines to $a_2=-10^{-11.523}$, and the dot--dashed
    lines to $a_2=-10^{-11.155}$.  The dotted
    lines show the corresponding results for a universe with a
    non--zero vacuum energy.}}
\label{STresult2}
\end{figure}

\subsubsection{Constraints on $\omega_0$ for a universe containing
    matter, radiation and nonzero vacuum energy}

We can repeat the previous analysis for the more realistic case of a
universe with a period of late--time vacuum domination.  We now need
the time of matter--radiation equality, $N_{eq1}$, and
the time of matter--vacuum equality, $N_{eq2}$.

Assuming spatial flatness, a large $\omega_0$, $h=0.7$,
$\rho_{\Lambda0}/\rho_{m0}=2.7$, and $\rho_{r0}$ as above gives
\begin{equation*}
N_{eq1} \simeq -8.06-\frac{1}{2}\beta(\psi_{eq1}-\psi_{\infty})^2,
\end{equation*}
and
\begin{equation*}
N_{eq2} \simeq -0.33-\frac{1}{2}\beta(\psi_{eq2}-\psi_{\infty})^2.
\end{equation*}

We can now evolve $\psi(N)-\psi_{\infty}$ from $N_{eq1}$ to $N_{eq2}$
using (\ref{solution}) and the same boundary conditions as before.
(In finding $N_{eq2}$ we used an iterative method to evaluate
$\psi_{eq2}-\psi_{\infty}$).  We then use (\ref{solution2})
to evolve the field from $N_{eq2}$ to $N_0$; the constants $C$ and
$D$ in (\ref{solution2}) are set by matching
$\psi(N)-\psi_{\infty}$ and its first derivative with (\ref{solution})
at $N_{eq2}$.  Finally, we can calculate $\omega_0$ using
(\ref{w(t)}).  The results of this procedure are shown in figure \ref{STresult2}.

\section{Discussion}

Using the framework provided by ST theories of gravity we have investigated
the effect of a time--varying $G$ during primordial nucleosynthesis.
We determined the effect on primordial nucleosynthesis numerically, using a
modified version of the Kawano code \cite{Kaw92}, and constrained the
parameters of the underlying theory using these results.  Our results
are consistent with the interpretation that the abundance
of $^4$He is primarily only sensitive to the value of $G$ at the time
when weak interactions freeze out, for the class of ST models studied.

Using our numerically determined constraints on the evolution of $G$, we
imposed the 2$\sigma$ upper and lower bounds shown in figure
\ref{BDresult} on the BD parameter $\omega$.  For
a constant $G$ (i.e. $a_2=0$) we get the bounds $\omega \gtrsim 332$ or
$\omega \lesssim -37$, for a universe containing matter and radiation
only, and the bounds  $\omega \gtrsim 277$ or $\omega \lesssim -31$,
for a universe containing matter, radiation and a nonzero vacuum
energy.  As the parameter $a_2$ is increased, the strength of the upper bound decreases
whilst that of the lower bound increases; the opposite behaviour occurs
as $a_2$ is decreased.  This is because the strength of the
bound on $\omega$ is essentially due to the allowed value of $G$ at
matter--radiation equality.  If primordial nucleosynthesis allows this value to be vastly
different from $G_N$ then a significant evolution of $G$ during matter
domination, and hence a low $\vert \omega \vert$, is permitted.  A value of
$G_{eq}$ close to $G_N$ means that only a very slow variation of $G$,
and hence large $\vert \omega \vert$ is permitted.  As $a_2>0$ corresponds to an
increasing $G$ during primordial nucleosynthesis this corresponds to a higher value
of $G_{eq}$ and hence a tighter upper bound on $\omega$ and a looser
lower bound (as $G$ decreases during matter domination for $\omega >-3/2$ and
increases for $\omega <-3/2$); $a_2<0$ corresponds to $G$ decreasing
during radiation domination, and so has the opposite effect.  

The more stringent effect of a nonzero value of
$a_2$ on the upper bounds is due to the current observational determinations of the
$^4$He abundance \cite{Bar03} disfavoring $G>G_N$ during primordial
nucleosynthesis (see e.g. \cite{Bar04}).

The interpretation of the constraints on the more general ST theories
is a little more complicated, due to the increased complexity of the
theories.  The bounds on $\omega(\phi)$ at matter--radiation equality,
shown in figure \ref{STresult1}, are seen to be stronger for smaller
$\beta$ and weaker for larger for $\beta$ (assuming $\beta$ to be
small enough to safely ignore the effect of the $e^-e^+$ kick
analysed by \cite{Dam99}).  This should be expected as $\omega \sim \beta^{-2}$ in the models
we are studying.  We see from the constraints on $\omega$
at the present day, shown in figure \ref{STresult2},
that the bounds on $\omega$ become tighter as $\beta$ gets
very small and large, with an apparent minimum in the
bound at $\beta \sim 0.2$.  The tight bounds for very small $\beta$
are due to the tight bounds on $\omega_{eq}$ for small $\beta$. The
tight bounds at large $\beta$ are due to the
attraction towards GR at late times that occurs for this class of ST
theories (see e.g. \cite{Dam93}).  This attractor mechanism is
more efficient for larger values of $\beta$, as can be seen from
(\ref{master}), and at late times so $\omega$ is drawn to a larger
value for a larger $\beta$.

The inclusion of a late--time vacuum--dominated stage of the Universe's
evolution is to weaken slightly the bounds that can be placed on
$\omega$ at the present day, as can be seen from figures
\ref{BDresult} and \ref{STresult2}.  This weakening of the bounds is due to
a shortening of the matter--dominated period of the Universe's history
which is essentially the only period, after the effects of the free
scalar--dominated phase become negligible, during which $G$ evolves.

We find that the constraints that can be imposed upon the present day
value of $\omega$ from primordial nucleosynthesis are, for most of the allowed parameter space, considerably
weaker than those obtained from observations within the solar system.
To date, the tightest constraint upon $\omega_0$ are imposed by
Bertotti, Iess and Tortora \cite{Ber03} who find $\vert \omega_0 \vert
\gtrsim 40000$, to $2\sigma$.  This constraint is obtained from observations of the
Shapiro delay of radio signals from the Cassini spacecraft as it
passes behind the Sun.  We consider
the constraints imposed upon $\omega$ here to be complementary to
these results as they probe different length and time scales, as well
as different epochs of the Universe's history.

\noindent\\

{\large \bf ACKNOWLEDGEMENTS}\\

T.C. is supported by PPARC.


\begin{thebibliography}{[]}
\bibitem{footnote} It has been suggested (see e.g. \cite{Asc04}) that these
  observations could be due to a different isotopic magnesium
  abundance.  However, it has also been shown that matching the isotopic abundances so as to
emulate the apparently observed variation in alpha also creates an over abundance of
nitrogen \cite{Fen05}.
\end{thebibliography}
\end{document}